\documentclass[]{style/ceurart}
\usepackage{placeins}
\usepackage{multirow}   
\usepackage{booktabs}   
\usepackage{listings}
\usepackage{graphicx}
\usepackage{amsmath}
\DeclareMathOperator*{\argmax}{arg\,max}
\usepackage{float}
\usepackage{tcolorbox}     
\usepackage{hyperref}      
\begin{document}

\copyrightyear{2026}
\copyrightclause{Copyright for this paper by its authors.
  Use permitted under Creative Commons License Attribution 4.0
  International (CC BY 4.0).}

\conference{CLEF 2026: Conference and Labs of the Evaluation Forum, September 21-24, 2026, Jena, Germany}

\title{Language-Routed RAG and Direct Option Scoring for Multilingual Financial QA: DS@GT at FinMMEval}
\title[mode=sub]{Notebook for the AI for Finance Lab at CLEF 2026 }

\author[1]{Justice Ayela}[
    orcid=0009-0001-3131-362X,
    email=jayela3@gatech.edu,
    url= https://www.linkedin.com/in/justiceayela/
]
\cormark[1]
\fnmark[1]

\author[1]{Kabir Sahni}[
    orcid=0009-0000-5795-5690,
    email=ksahni30@gatech.edu,
    url= 
    https://www.linkedin.com/in/kabir-sahni-8a500922a/
]
\cormark[0]
\fnmark[1]

\address[1]{Georgia Institute of Technology, North Ave NW, Atlanta, GA 30332}
\cortext[1]{Corresponding author.}

\begin{abstract}
We present DS@GT's submission to FinMMEval 2026 Task~1, a multilingual
financial exam question answering benchmark spanning English, Spanish, Greek,
Chinese, and Hindi.
Financial certification exams such as the CFA, EFPA, and CPA demand
structured domain reasoning that standard NLP benchmarks do not capture, and
this challenge compounds across languages where retrieval and representation
infrastructure is underdeveloped.
We build a retrieval-augmented pipeline on LangGraph that detects query
language, retrieves semantically relevant exemplars from a 30{,}209-entry
multilingual knowledge base using BGE-M3 embeddings and FAISS indexing. The system then
scores answers via Retrieval-Augmented Direct Scoring (RADS)~---~reading
next-token log-probabilities over candidate option letters rather than
generating free-form output.
For low-resource languages, we fuse per-language and cross-lingual retrieval
indices using weighted Reciprocal Rank Fusion~\cite{cormack2009rrf}.
Model selection is language-routed: Qwen3-14B~\cite{qwen3} for Arabic,
Chinese, and Hindi; Qwen2.5-14B~\cite{qwen25} for English; and
Llama-3.1-8B~\cite{llama3} for Greek~---~a routing derived from empirical
ablations that reveal substantial language-asymmetric performance gaps.
Notably, chain-of-thought prompting significantly degrades Greek accuracy
($90.7\%\rightarrow 20.9\%$), and enabling Qwen3's default thinking mode
collapses Arabic RADS performance to near-chance levels.
Our results indicate that effective multilingual financial reasoning requires
language-aware retrieval, model routing, and deliberate scoring strategy
selection.
\end{abstract}


\maketitle

\section{Introduction}

Professional financial certification exams~---~CFA, EFPA, CPA, and their
equivalents~---~are not general knowledge tests.
They require candidates to move between regulatory frameworks, apply
domain-specific accounting logic, and reason through multi-step valuation
problems under time pressure.
This is a qualitatively different challenge from the factual recall tasks on
which most large language models are benchmarked, exposing a real gap: high
performance on standard NLP benchmarks does not reliably predict competence
on structured financial reasoning.
 
The problem compounds across languages.
Most financial NLP research is conducted in English, yet financial
professionals operate in Spanish, Greek, Mandarin, Hindi, and dozens of other
languages.
Regulatory context shifts with jurisdiction, terminology does not translate
cleanly, and low-resource languages lack the pretraining data density that
English-centric models rely on.
A system achieving 80\% accuracy on CFA-style English questions may perform substantially worse on equivalent Greek or Hindi questions due to weaker retrieval and representation infrastructure.
 
FinMMEval 2026 Task~1 provides a concrete setting to study this: five
datasets spanning English, Spanish, Greek, Chinese, and Hindi, covering
domains from investment regulation to corporate accounting, with over 1{,}600
multiple-choice questions drawn from actual certification exam
material~\cite{FinMMEval2026}.
 
\begin{figure}
    \centering
    \includegraphics[width=1\linewidth]{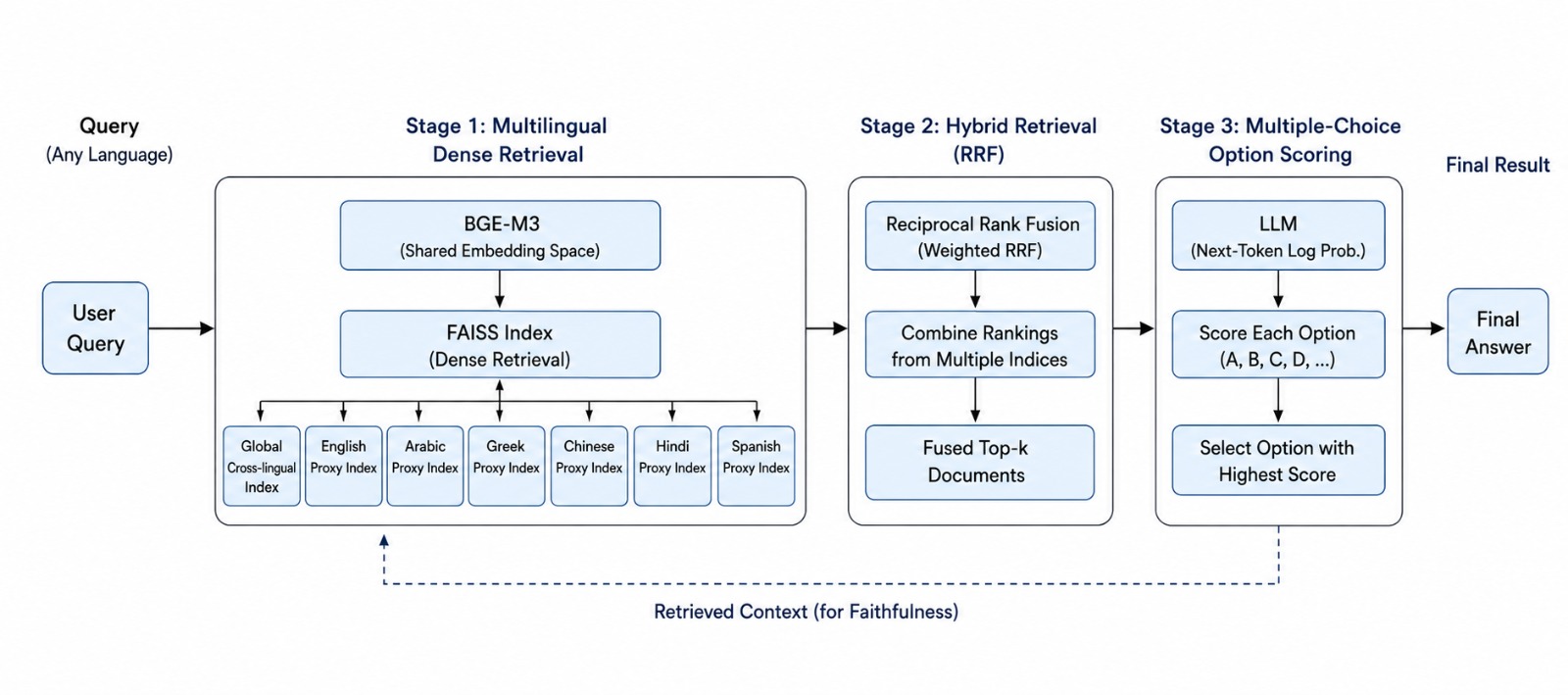}
    \caption{The three-stage multilingual hybrid retrieval and multiple-choice reasoning architecture}
    \label{fig:llm_architecture}
\end{figure}

We treat this as a retrieval-augmented reasoning problem. Our system, built on a LangGraph orchestration pipeline, routes each question
through language detection and concept extraction, retrieves semantically
relevant exemplars from a multilingual FAISS index built with BGE-M3
embeddings \cite{chen2024m3embedding}, fuses per-language and cross-lingual retrieval
signals using weighted Reciprocal Rank Fusion~\cite{cormack2009rrf}, and
scores answers via next-token log-probability rather than free-form
generation, as shown in Figure \ref{fig:llm_architecture} .
The architecture handles out-of-distribution languages by falling back to
cross-lingual retrieval.
 
The contributions of this work are:
\begin{enumerate}
    \item A structured multilingual RAG pipeline tailored to financial exam
          reasoning.
    \item A two-layer exemplar retrieval strategy combining per-dataset and
          global indices with reciprocal rank fusion.
    \item An empirical analysis of cross-lingual performance across six
          financially distinct corpora, including previously unreported
          language-asymmetric effects of reasoning strategies and model
          selection.
\end{enumerate}

Our findings reveal that effective multilingual financial reasoning requires language-aware design at every stage of the pipeline.
Retrieval-Augmented Direct Scoring (RADS) ties or exceeds few-shot generation across every language tested while eliminating parse failures entirely. Language-routed model selection is essential: no
single model is best across all languages — Llama-3.1-8B outperforms
all Qwen variants on Greek by $+19.5$ percentage points, while
Qwen3-14B unexpectedly regresses by $-22.9$ percentage points on
English relative to Qwen2.5-14B despite improving every other language
we tested. Chain-of-thought prompting with self-consistency proves
sharply language-asymmetric, significantly degrading Greek
classification ($90.7\% \rightarrow 20.9\%$) and Hindi ($-19$~pp)
while only marginally helping Chinese. Finally, a $17$ percentage-point
gap between our best open-weight system and Claude Opus 4.7 on Hindi
indicates that base-model capability — not retrieval quality — is the
dominant remaining bottleneck for reproducible multilingual financial
QA. The implementation can be found at
\url{https://github.com/dsgt-arc/finmmevaltask1-2026}.

\section{Related Work}

\subsection{Retrieval-Augmented Generation (RAG)}
Retrieval-augmented generation (RAG) has emerged as a pivotal framework in natural language processing, enhancing the performance of large language models on knowledge-intensive tasks. The
seminal work by Lewis et al.~\cite{lewis2020rag} introduced this
paradigm, demonstrating how combining retrieval-based and
generation-based approaches can improve accuracy by grounding model outputs in external documents. Subsequent work by Izacard et
al.~\cite{izacard2021leveraging} further refined RAG by introducing
Fusion-in-Decoder (FiD), which enhances generation by aggregating
information from multiple retrieved documents. We adopt RAG at the
\emph{exemplar} level rather than the document level. This shift lets us pair retrieval with
option-letter scoring rather than text generation
(Section~\ref{sec:rads}), making the predicted answer a deterministic argmax over next-token log-probabilities.

\subsection{Multilingual Dense Retrieval}
For cross-lingual exemplar retrieval, we use BGE-M3 \cite{chen2024m3embedding}, a multilingual embedding model that supports dense retrieval across 100+ languages within a unified vector space. Vector indexing is performed with FAISS \cite{douze2024faiss,johnson2019billion}, which provides efficient nearest-neighbor search at scale via inverted-file and product-quantization indices. Unlike traditional bi-encoder pipelines that require separate per-language models, BGE-M3's shared embedding space enables zero-shot cross-lingual retrieval used for low-resource target languages.

\subsection{Reciprocal Rank Fusion for Hybrid Retrieval}
For low-resource languages, we combine multiple retrieval indices using Reciprocal Rank Fusion (RRF) introduced by Cormack et al. \cite{cormack2009rrf}. RRF simply sorts the documents according to a naive scoring formula, aggregating rankings as $\text{RRFscore}(d) = \sum_{r \in R} \frac{1}{k + r(d)}$. We extend this with per-retriever weights to favor native-language indices over cross-lingual proxies: $\text{RRFscore}(d) = \sum_i \frac{w_i}{k + r_i(d)}$, where $r_i(d)$ is the rank of document $d$ in retriever $i$ and $w_i$ is its trust weight (we use $w=1.0$ for the global cross-lingual index and $w=2.0$ for per-language indices). This weighted variant is well-suited to combining a global cross-lingual index with weighted per-language proxy indices.

\subsection{Multiple-Choice QA via Option Scoring}
Rather than generating an answer letter via free-form decoding, we score each option by the model's next-token log-probability following the methodology established by Hendrycks et al. ~\cite{hendrycks2021mmlu} in MMLU. This has become the standard for LLM evaluation on more recent benchmarks like BizBench \cite{koncelkedziorski2024bizbench}, FinBen \cite{xie2024finben}, and PIXIU \cite{xie2023pixiu} because it eliminates parse failures, produces deterministic predictions, and removes the formatting bias inherent in generation-based scoring. Brown et al. \cite{brown2020gpt3} originally popularized log-likelihood-based few-shot evaluation in the GPT-3 paper, and the EleutherAI lm-evaluation-harness \cite{eval-harness} has standardized the implementation.

\subsection{Multi-Solver Aggregation and Confidence Calibration}
When combining heterogeneous solvers (option scoring vs. few-shot generation vs. CoT), we weight each solver's vote by a product of three signals: faithfulness, calibrated confidence, and terminology overlap. The faithfulness component draws on the framing in Angulo and Yeste \cite{angulo2025agentic}, where retrieval quality and entity preservation are treated as core signals for grounding generated outputs in retrieved context. For confidence, we use adaptive temperature scaling derived from Guo et al. \cite{guo2017calibration}, who demonstrated that modern neural networks are systematically overconfident and that softmax-temperature calibration reduces the gap between confidence and accuracy without changing the argmax prediction.

\section{Methodology}

\subsection{Baseline Methods}
For retrieval, we use the BGE-M3 model \cite{chen2024m3embedding} to embed both the query and the exemplar knowledge base into a shared 1024-dimensional multilingual vector space. Exemplars are indexed in FAISS~\cite{johnson2019billion} using cosine similarity over L2-normalized embeddings, defined by:
\begin{equation}
\cos(\mathbf{x}, \mathbf{y}) = \frac{\langle \mathbf{x}, \mathbf{y} \rangle}{\|\mathbf{x}\| \, \|\mathbf{y}\|},
\end{equation}
where $\mathbf{x}, \mathbf{y} \in \mathbb{R}^{1024}$ are the query and exemplar embeddings produced by BGE-M3, $\langle \mathbf{x}, \mathbf{y} \rangle = \sum_{i=1}^{1024} x_i y_i$ is their inner product, and $\|\mathbf{x}\| = \sqrt{\langle \mathbf{x}, \mathbf{x} \rangle}$ is the Euclidean norm. Given all embeddings are L2-normalized at index time (i.e.\ $\|\mathbf{x}\| = \|\mathbf{y}\| = 1$), cosine similarity reduces to the inner product, this allows efficient search without an explicit normalization step at query time. All experiments were executed on the Georgia Institute of Technology Phoenix Research Computing Cluster [1] GPU nodes.

\subsection{Knowledge Base Construction}

We construct a multilingual exemplar knowledge base (KB) $\mathcal{E} = \{(q_i, c_i, a_i, \ell_i)\}_{i=1}^{N}$ where each entry consists of a question $q_i$, choices $c_i$, gold answer $a_i \in \{A, B, C, D\}$, and language label $\ell_i$. The knowledge base aggregates the datasets that have been made publicly available through the FinMMEval Hugging Face collection; the original split names on the dataset cards do not restrict usage, and they were reorganized as needed during training as per policy. We aggregated the monolingual training and development sets from each language into a unified multilingual training set at inference time. The dev-test sets were similarly merged to form a multilingual development set. Additionally, $8{,}877$ single-choice standardized exemplars were used from the CFinBench \cite{nie2024cfinbench} dataset. The distribution of data samples across these collections is summarized in Table~\ref{tab:kb}. 

The knowledge base contains $N = 30{,}209$ exemplars. Each entry is rendered into a canonical \texttt{QUESTION: \ldots OPTIONS: (A) \ldots} format that matches the input format of target queries at inference time. 

\begin{table}[htbp]
\caption{Distribution of data samples aggregated across Knowledge Base}
\label{tab:kb}
\centering
\small
\setlength{\tabcolsep}{8pt}
\renewcommand{\arraystretch}{1.15}
\begin{tabular}{@{}llrrr@{}}
\toprule
\textbf{Language} & \textbf{Dataset} & \textbf{Train} & \textbf{Val} & \textbf{Test} \\
\midrule
\multicolumn{5}{@{}l}{\textit{Subtask 1: Financial Exam Question Answering}} \\
\midrule
English  & BhashaBench-Finance              & --   & --    & 13{,}451 \\
         & FinMMEval-CFA-CPA                & 449  & --    & --       \\
\midrule
Greek    & Plutus-MultiFin                  & 171  & 43    & 54       \\
\midrule
Spanish  & TheFinAI/FLARE-ES-MultiFin       & --   & --    & 230      \\
\midrule
Arabic   & SahmBench/Arabic-Business-MCQ    & 274  & --    & 183      \\
         & SahmBench/Arabic-Accounting-MCQ  & 249  & --    & 167      \\
\midrule
Chinese  & CFinBench (single\_choice)       & --   & 8{,}877 & --     \\
         & FinMMEval-CFA-CPA                & 27   & --    & --       \\
\midrule
Hindi    & BhashaBench-Finance              & --   & --    & 5{,}982  \\
\bottomrule
\end{tabular}
\end{table}

\subsection{Language-Aware Index Partitioning}

To enforce stylistic and lexical consistency between target queries and retrieved exemplars, we partition the KB by language and construct separate FAISS shards $\{I_\ell\}_{\ell \in \mathcal{L}}$ for each language $\ell \in \{$ar, el, zh, en, hi, es$\}$, in addition to a global cross-lingual index $I_\text{global}$ containing all $N$ exemplars. This partitioning allows the retriever to enforce language consistency when sufficient native exemplars exist and to fall back to cross-lingual signal otherwise — a routing decision formalized in Section~\ref{sec:firststage}.

All shards share the same 1024-dimensional BGE-M3 embedding space and use \texttt{IndexFlatIP}, so cosine scores are directly comparable across shards. The largest shards are English ($|I_\text{en}| = 14{,}057$) and Chinese ($|I_\text{zh}| = 8{,}896$); the smallest is Greek ($|I_\text{el}| = 171$). Spanish ($|I_\text{es}| = 230$) and Hindi ($|I_\text{hi}| = 5{,}982$) sit between these extremes. Our analysis indicates significant imbalance in exemplar distribution across language shards.

\subsection{First-Stage Retrieval}
\label{sec:firststage}

\paragraph{Adaptive Multi-Scope Routing.} Given a target question $q$ with detected language $\ell$, the retriever selects one of three search strategies based on per-language coverage:

\begin{itemize}
    \item \textbf{Dedicated per-language search.} When $|I_\ell| \geq \tau$ ($\tau = 20$), we strictly search the per-language shard $I_\ell$. This strict-isolation regime is used for Arabic ($|I_\text{ar}|$), Chinese ($|I_\text{zh}|$), English ($|I_\text{en}|$), Hindi ($|I_\text{hi}|$), and Spanish ($|I_\text{es}|$) in our final configuration.
    \item \textbf{Weighted RRF.} When $0 < |I_\ell| < \tau$, we expand the search to combine $I_\text{global}$ and $I_\ell$ via a weighted variant of Reciprocal Rank Fusion \cite{cormack2009rrf} with per-retriever trust weights: 
    \begin{equation}
        \text{RRFscore}(d) = \sum_{i \in \{\text{global},\, \ell\}} \frac{w_i}{k + r_i(d)},
        \label{eq:rrf}
    \end{equation}
    where $r_i(d)$ is the rank of exemplar $d$ in retriever $i$, $k = 60$ is the RRF damping constant, and $w_\text{global} = 1.0$, $w_\ell = 2.0$ favors native-language signal over cross-lingual signal when both are available.
    \item \textbf{Global cross-lingual.} When $|I_\ell| = 0$ the query is in a language absent from the knowledge base or the detected language is unknown, we search $I_\text{global}$ alone.
\end{itemize}

\paragraph{Exemplar Selection.} From each strategy we retain the top-$k$ exemplars ($k = 5$) ranked by the chosen scoring function (raw cosine similarity for dedicated/global search, RRF score for fused search). Retrieval is the only step that uses the embedding model; downstream answer scoring sees the retrieved exemplars as plain text and does not re-encode them.

\paragraph{Language Detection.} For queries in the final test suite, language is detected by mapping the competition-issued ID prefix (e.g.\ \texttt{ar-task1-final-test-001} $\rightarrow$ ar). For inputs without language metadata, we fall back to a Unicode script detection. 

\subsection{Language-Routed Answer Scoring}
\label{sec:rads}

\paragraph{Prompt Construction.}
Given a target query $q$ and its top-$k=5$ retrieved exemplars $\{e_1, \ldots, e_k\}$, we carefully designed a prompt with a system instruction that fixes the model's role as a financial exam solver and constrains its response to a single answer letter, and a user message that interleaves the $k$ solved exemplars with the target query. The models used are Qwen3-14B~\cite{qwen3}, Qwen2.5-14B-Instruct~\cite{qwen25}, and Meta-Llama-3.1-8B-Instruct~\cite{llama3}, selected per-language via the routing in Equation~\ref{eq:routing}. The full prompt template can be found in Appendix~\ref{app:rads-prompt}.

\paragraph{Retrieval-Augmented Direct Scoring (RADS).}
We perform a single forward pass over the prompt and read the next-token log-probability of each candidate option letter $\ell \in \mathcal{L}_q \subseteq \{A, B, C, D, E\}$ (where $\mathcal{L}_q$ is the set of valid option letters for question $q$). Letting $f_\theta(\cdot)$ denote the model's logits at the final prompt position and $t_\ell$ the tokeniser ID of letter $\ell$, the predicted answer is:
\begin{equation}
\hat{a} = \argmax_{\ell \in \mathcal{L}_q} \, f_\theta(\text{prompt})[t_\ell].
\label{eq:rads}
\end{equation}
The scoring methodology popularized by MMLU~\cite{hendrycks2021mmlu} and EleutherAI \texttt{lm-evaluation-harness}, with two adaptations for our setting: (i) only the valid letters $\mathcal{L}_q$ for the specific question are scored, so 3-option items do not waste probability mass on a non-existent ``D''; and (ii) the prompt is augmented with retrieved exemplars rather than fixed templates, making the score retrieval-augmented in generation without the parse-failure surface.

\paragraph{Language-Routed Model Selection.}
The underlying LLM $\theta$ is selected per query language $\ell$ across four model families (Qwen2.5-14B, Qwen3-14B, Llama-3.1-8B, Qwen2.5-7B baseline) on language-specific dev sets. The final routing is:
\begin{equation}
\theta(\ell) =
\begin{cases}
\text{Qwen3-14B} & \ell \in \{\text{ar}, \text{zh}, \text{hi}\} \\
\text{Qwen2.5-14B} & \ell = \text{en} \\
\text{Llama-3.1-8B} & \ell = \text{el} \\
\text{Qwen3-14B} & \text{otherwise (default)}
\end{cases}
\label{eq:routing}
\end{equation}

\paragraph{Implementation Note: Disabling Qwen3 Thinking Mode.}
Qwen3 models default to emitting \texttt{<think>\ldots</think>} reasoning tokens before the answer~\cite{qwen3}. Under RADS, these reasoning tokens displace the answer letter from the next-token position and break Equation~\ref{eq:rads}. We disable thinking mode by passing \texttt{enable\_thinking=False} to the chat-template constructor, which restores the answer letter to the immediate next-token position without altering the model weights. This step is critical: with thinking enabled, RADS accuracy on Arabic dropped from 71.7\% to near-chance levels in our preliminary experiments.

\section{Results}
\label{sec:results}

We report three sets of experiments. Section~\ref{sec:routing-results} reports the adaptive multi-scope retriever and quantifies the effect of expanding the KB to add native Hindi exemplars. Section~\ref{sec:rads-results} compares RADS option scoring against generation, chain-of-thought, and translation-based solvers. Section~\ref{sec:model-results} reports the routing of the model per-language and contrasts our local system with the API baselines of the frontier. All dev-leaderboard scores are exact-match accuracy on the official FinMMEval dev set.

\subsection{Adaptive Multi-Scope Routing Results}
\label{sec:routing-results}

We rebuilt the knowledge base in two configurations to isolate the contribution of native-language coverage:

\begin{itemize}
    \item \textbf{KB v1} (5 datasets, 6{,}996 exemplars): Arabic train splits, Greek Plutus-MultiFin train, CFA-aligned MMLU validation, CFinBench \texttt{single\_choice/dev}, and the bilingual FinMMEval-CFA-CPA. Hindi has no native exemplars and is served via three-way RRF over the global index, an English proxy index, and an Arabic proxy index.
    \item \textbf{KB v2} (10 datasets, 30{,}209 exemplars): KB v1 + Arabic test splits + BhashaBench-Finance (English subset + Hindi subset) + TheFinAI/FLARE-ES-MultiFin + CFinBench \texttt{single\_choice/val}. Both Hindi and Spanish now have dedicated per-language indices and are served from those rather than via cross-lingual proxies.
\end{itemize}

\begin{figure}[htbp]
    \centering
    \includegraphics[width=1\linewidth]{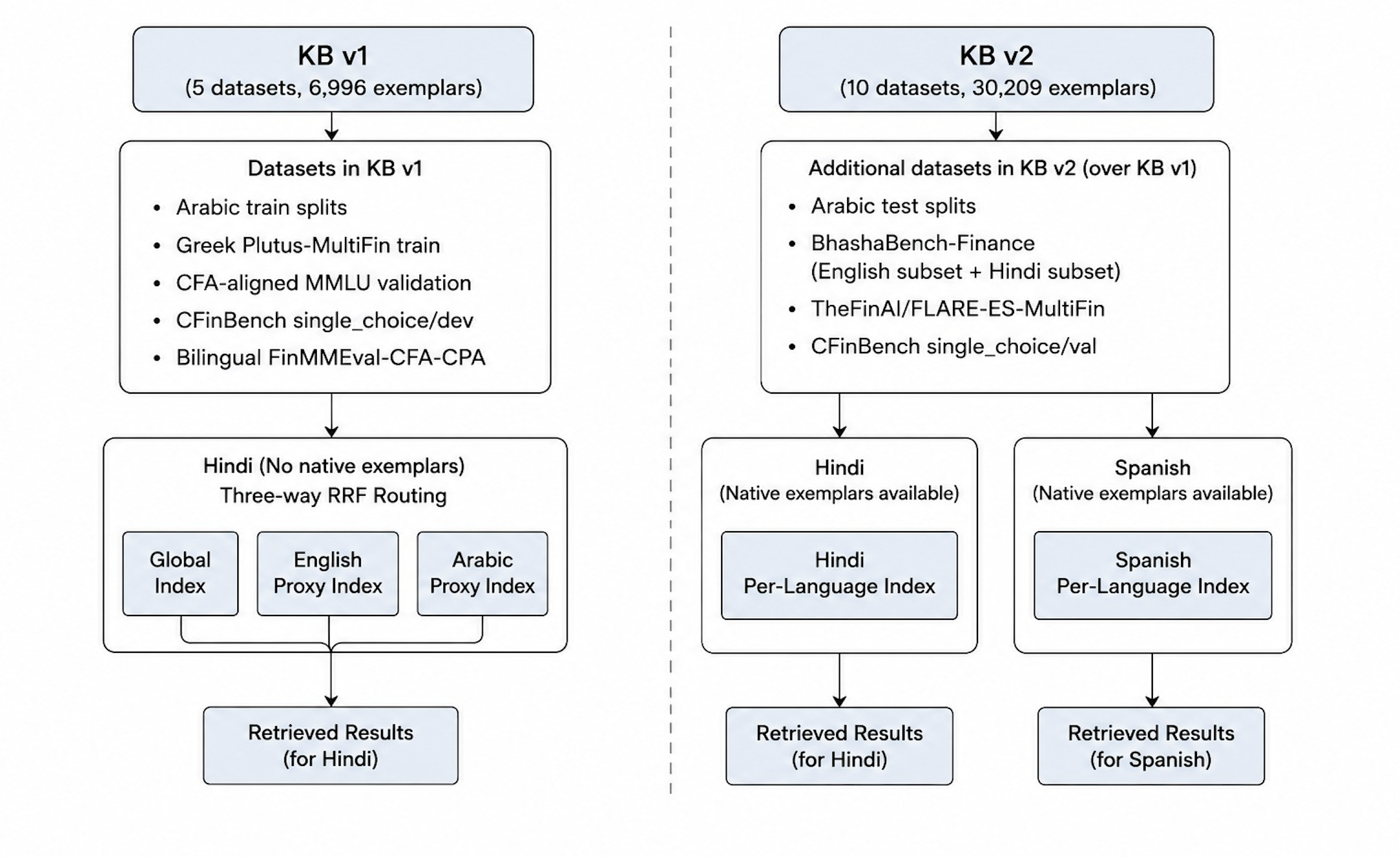}
    \caption{Adaptive Multi-Scope Routing Results}
    \label{fig:adap_multi}
\end{figure}

Figure~\ref{fig:adap_multi} explains the adaptive multi-scope routing results comprehensively. Table~\ref{tab:kb-versions} reports per-language exemplar counts, and Table~\ref{tab:routing-dev} compares dev-leaderboard accuracy under the two KBs with all other system components held constant (BGE-M3 retrieval, RADS scoring, language-routed model selection).

\begin{table}[htbp]
\caption{Per-language exemplar counts in the two KB configurations. KB v2 adds native Hindi and Spanish coverage (previously absent) and expands the Chinese and English shards.}
\label{tab:kb-versions}
\centering
\small
\begin{tabular}{@{}lrrr@{}}
\toprule
\textbf{Language} & \textbf{KB v1} & \textbf{KB v2} & \textbf{Routing in v2} \\
\midrule
Arabic (ar)  & 523    & 873      & dedicated \\
Greek (el)   & 171    & 171      & dedicated \\
Chinese (zh) & 156    & 8{,}896  & dedicated \\
English (en) & 606    & 14{,}057 & dedicated \\
Hindi (hi)   & 0      & 5{,}982  & dedicated (was 3-way RRF in v1) \\
Spanish (es) & 0      & 230      & dedicated (was global in v1) \\
\midrule
\textbf{Total} & \textbf{1{,}456} & \textbf{30{,}209} & \\
\bottomrule
\end{tabular}
\end{table}

\begin{table}[!htb]
\caption{Dev-leaderboard accuracy under KB v1 vs.\ KB v2. All language-specific scores hold constant or improve under v2; no regressions are observed.}
\label{tab:routing-dev}
\centering
\small
\begin{tabular}{@{}lrrr@{}}
\toprule
\textbf{Language} & \textbf{KB v1} & \textbf{KB v2} & \textbf{$\Delta$} \\
\midrule
Arabic   & 77.32\% & \textbf{78.00\%} & $+0.68$ \\
Chinese  & 69.00\% & \textbf{70.00\%} & $+1.00$ \\
English  & 75.71\% & 75.71\% & $-$ \\
Hindi    & 78.00\% & \textbf{79.00\%} & $+1.00$ \\
\midrule
Average  & 75.01\% & \textbf{75.68\%} & $+0.67$ \\
\bottomrule
\end{tabular}
\end{table}

\FloatBarrier
The observation follows:

\paragraph{Native exemplars outperform cross-lingual proxies on Hindi.}
The Hindi dev-leaderboard score improved from $78.0\%$ (v1, three-way RRF over global, English, Arabic) to $79.0\%$ (v2, dedicated Hindi shard of $5{,}982$ exemplars). The gain is modest because the underlying Qwen3-14B already handled the cross-lingual proxies well with RRF over English and Arabic exemplars, which supplied relevant signal even without native Hindi text. Section~\ref{sec:model-results} reports a much larger gap when this is compared against frontier API models.

\subsection{Retrieval-Augmented Direct Scoring (RADS) Results}
\label{sec:rads-results}

We compare four solver families on internal validation splits, holding retrieval and language routing fixed:

\begin{itemize}
    \item \textbf{Solver A (Few-shot RAG)} — Predicts an answer letter by greedy decoding after a 5-shot prompt.
    \item \textbf{Solver B (RADS)} — Retrieval-augmented direct scoring (this paper, Equation~\ref{eq:rads}).
    \item \textbf{Solver C (NLLB-200 translation + English RADS)} — Translates the source-language query into English, then solves with English exemplars.
    \item \textbf{Solver D (CoT + Self-Consistency)} — Chain-of-thought with self-consistency ($N{=}5$, $T{=}0.7$).
\end{itemize}

\begin{figure}[htbp]
\centering
\includegraphics[width=\columnwidth]{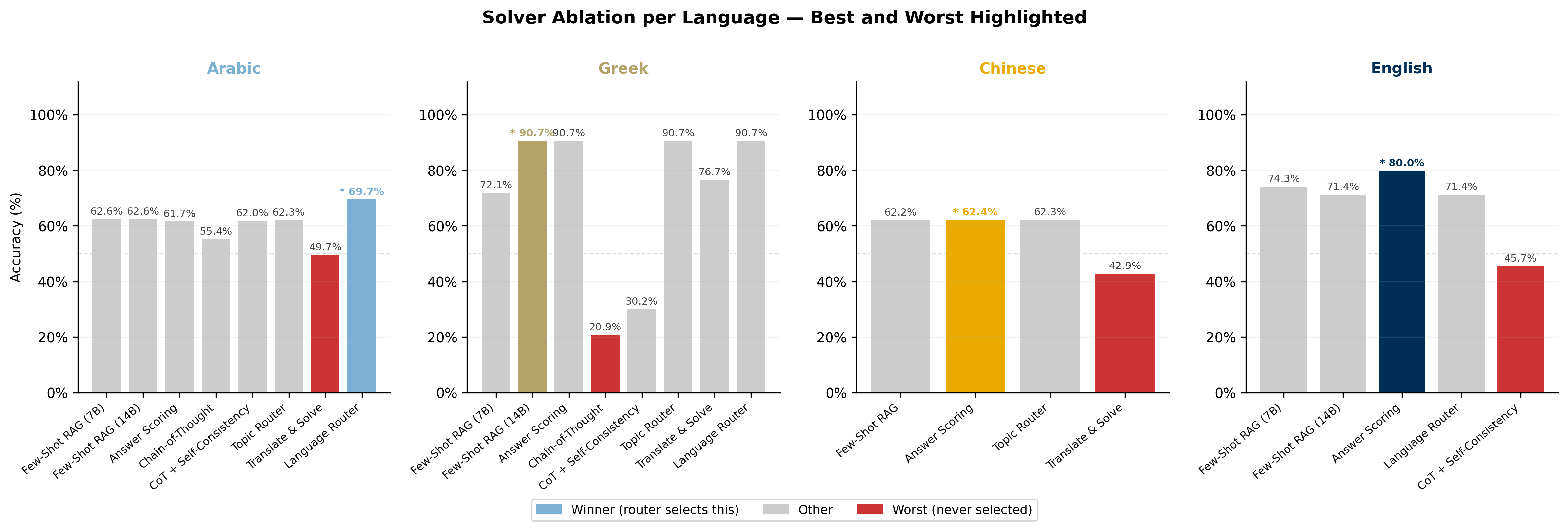}
\caption{Per-language accuracy of the four solver families under Qwen2.5-14B (except Greek, which uses Llama-3.1-8B). Solver B (RADS) ties or exceeds Solver A on every language except Arabic with Qwen2.5; the Arabic gap is closed by upgrading to Qwen3-14B (see Table~\ref{tab:model-routing}). Solver C (NLLB translation) is the worst performer in every setting.}
\label{fig:solver-ablation}
\end{figure}

Table~\ref{tab:solver-ablation} reports per-language accuracy. Figure~\ref{fig:solver-ablation} visualises the same results.

\begin{table}[htbp]
\caption{Solver ablation by language. Accuracies measured on internal validation splits with the underlying language-routed model (Qwen2.5 base for solver comparison; Qwen3 columns reported separately for the final routing). Bold marks the best solver per language under a fixed base model.}
\label{tab:solver-ablation}
\centering
\small
\setlength{\tabcolsep}{4pt}
\begin{tabular}{@{}lrrrr@{}}
\toprule
\textbf{Language ($N$)} & \textbf{Solver A} & \textbf{Solver B} & \textbf{Solver C} & \textbf{Solver D+SC} \\
\midrule
Greek el (43)        & \textbf{90.7\%} & \textbf{90.7\%} & 76.7\% & 20.9\% \\
Chinese zh (8{,}869) & 62.2\%          & \textbf{62.4\%} & 42.9\% & 66.3\%$^\dagger$ \\
English en (35)      & 71.4\%          & \textbf{80.0\%} & --     & 45.7\% \\
Arabic ar (175)      & \textbf{69.7\%} & 61.7\%          & 49.7\% & 62.0\% \\
\bottomrule
\end{tabular}
\\[2pt]
\footnotesize{$^\dagger$ Chinese Solver D+SC was discontinued at $\approx 63\%$ completion ($5{,}580/8{,}869$ records) when rolling accuracy plateaued below Qwen3 Solver B; the value reported is the rolling estimate at stop time.}
\end{table}

Four findings follow:

\paragraph{RADS eliminates parse failures.}
Solver B achieves a $100\%$ parse rate across all $8{,}869$ Chinese validation questions because it never decodes free text — the answer is the argmax over four log probabilities. Solver A required a fallback parsing chain (regex on first letter, last letter, then a constrained single-token regeneration) and still produced ambiguous outputs on $\approx 2\%$ of Arabic queries in our preliminary runs.

\paragraph{Translate-then-solve is a strict regression.}
Solver C lags every other solver in every language we tested: $-13.9$~pp on Greek, $-19.5$~pp on Chinese, $-20.0$~pp on Arabic relative to the best solver per language. NLLB-200 translation discards numerical and named-entity precision that the financial MCQ task depends on. We rule out cross-lingual translation as a viable solver family for this domain.

\paragraph{CoT+SC is language-asymmetric.}
Solver D+SC degrades Greek significantly ($90.7\% \rightarrow 20.9\%$, $-69.8$~pp). Greek Plutus-MultiFin is text classification into six financial categories, and chain-of-thought prompting biases the model toward narrative completion rather than categorical assignment. On English MMLU-financial, D+SC drops from Solver B's $80.0\%$ to $45.7\%$ ($-34.3$~pp). The only positive signal is on Chinese, where preliminary D+SC tracking reached $\sim 66\%$ — comparable to Qwen3 Solver B at $66.6\%$ but $\sim 5\times$ slower per query.

\paragraph{Solver A wins on Arabic under Qwen2.5 but loses under Qwen3.}
Under Qwen2.5-14B, Solver A on Arabic (\textbf{69.7\%}) beats Solver B (\textbf{61.7\%}) by $+8$~pp. We initially exploited this with a weighted A+B aggregator (faithfulness $\times$ confidence $\times$ terminology overlap) reported at $69.7\%$. Upgrading to Qwen3-14B equalized the two solvers at $71.7\%$ on Arabic, reducing the aggregator's tie-break workload to near-zero. Section~\ref{sec:model-results} formalises this finding.

\subsection{Language-Routed Model Selection Results}
\label{sec:model-results}

We experimented with three open-weight LLM families across all five evaluation languages: Qwen2.5-14B-Instruct, Qwen3-14B, and Llama-3.1-8B-Instruct. Table~\ref{tab:model-routing} reports per-language accuracy under each model with Solver B (RADS) and otherwise identical retrieval/prompt construction.

\begin{table}[htbp]
\caption{Per-language model ablation under RADS. Bold marks the selected model per language in the final routing (Equation~\ref{eq:routing}); ``--'' indicates the combination was not run because preliminary results made it dominated.}
\label{tab:model-routing}
\centering
\small
\setlength{\tabcolsep}{6pt}
\begin{tabular}{@{}lrrr@{}}
\toprule
\textbf{Language} & \textbf{Qwen2.5-14B} & \textbf{Qwen3-14B} & \textbf{Llama-3.1-8B} \\
\midrule
Arabic (ar)  & 69.7\% & \textbf{71.7\%} & -- \\
Greek (el)   & 69.8\% & 71.2\%          & \textbf{90.7\%} \\
Chinese (zh) & 62.4\% & \textbf{66.6\%} & -- \\
English (en) & \textbf{80.0\%} & 57.1\% & -- \\
Hindi (hi)   & 72\%   & \textbf{78\%}   & -- \\
\bottomrule
\end{tabular}
\end{table}

\paragraph{Qwen3-14B regresses on English by $-22.9$~pp.}
Qwen3-14B scores $57.1\%$ on the English CFA-aligned MMLU dev set versus Qwen2.5-14B's $80.0\%$ on the same 35 questions. The regression replicates with the same prompt, exemplars, and RADS scoring. We attribute this to differences in instruction-tuning data distribution rather than to capability, since Qwen3 dominates Qwen2.5 on every non-English language we tested. The English routing in Equation~\ref{eq:routing} reverts to Qwen2.5-14B accordingly.

\paragraph{Llama-3.1-8B dominates Greek by $+19.5$~pp.}
Llama-3.1-8B-Instruct achieves $90.7\%$ on Greek Plutus-MultiFin validation versus Qwen3-14B's $71.2\%$. The gap aligns with Llama's stronger reported multilingual coverage on European languages; we use it for Greek regardless of the Qwen-first default.

\paragraph{Qwen3 is the right default for ar/zh/hi.}
Qwen3 improves Arabic by $+2.0$~pp, Chinese by $+4.2$~pp, and Hindi by $+6$~pp relative to Qwen2.5. Disabling thinking-mode tokens (\texttt{enable\_thinking=False}) was necessary to make RADS work.

\paragraph{The frontier-API ceiling remains $>15$~pp above our best local system on Hindi.}
Table~\ref{tab:hindi-frontier} compares our best local Hindi result against frontier API models on the same 100-question dev-leaderboard set. The 17~pp gap between Claude Opus 4.7 and Qwen3-14B is not closeable by retrieval improvements alone — KB expansion from cross-lingual proxies to native Hindi exemplars (Section~\ref{sec:routing-results}) added $+1$~pp, less than $6\%$ of the gap. We submit Qwen3-14B as the reproducible local system and report Claude as the API oracle upper bound.

\begin{table}[htbp]
\caption{Hindi dev-leaderboard accuracy across local and frontier models, all using the same retriever and prompt. The local system reaches $79\%$ on KB v2; the frontier API ceiling sits at $95\%$.}
\label{tab:hindi-frontier}
\centering
\small
\begin{tabular}{@{}lr@{}}
\toprule
\textbf{Model} & \textbf{Hindi dev accuracy} \\
\midrule
Qwen2.5-14B + RADS (KB v1, RRF proxy)    & 72\% \\
Qwen3-14B + RADS (KB v1, RRF proxy)      & 78\% \\
Qwen3-14B + RADS (KB v2, native shard)   & \textbf{79\%} \\
GPT-4o (few-shot, KB v1)                  & 85\% \\
Claude Opus 4.7 (few-shot, KB v1)         & \textbf{95\%} \\
\bottomrule
\end{tabular}
\end{table}

\subsection{End-to-End System Results}
\label{sec:e2e-results}

We submitted our final system to the FinMMEval Task 1 Financial Exam Question Answering 2026 challenge. Table~\ref{tab:final-leaderboard} reports our team's standing within the organizer-side final test results. Our system attained accuracies of $76.0\%$ on Arabic ($10$th), $73.5\%$ on Hindi ($7$th), $69.5\%$ on English ($9$th), and $64.5\%$ on Chinese ($9$th).

\begin{table}[!htb]
\caption{Final leaderboard results on the FinMMEval Task 1 test set ($N=200$ per language, $100\%$ coverage). For each language we report the top-ranked team, a representative mid-pack team, our system (DS@GT --- FinMMEval Task 1), and the lowest-ranked team. Accuracy is exact-match.}
\label{tab:final-leaderboard}
\centering
\small
\setlength{\tabcolsep}{6pt}
\begin{tabular}{@{}llrr@{}}
\toprule
\textbf{Language} & \textbf{Team} & \textbf{Rank} & \textbf{Accuracy} \\
\midrule
\multirow{4}{*}{Arabic}
& PooRi                              & 1st  & 97.5\% \\
& UyoAngle                           & 6th  & 86.5\% \\
& \textbf{DS@GT --- FinMMEval Task 1} & \textbf{10th} & \textbf{76.0\%} \\
& FinEval                            & 11th & 9.5\%  \\
\midrule
\multirow{4}{*}{Chinese}
& PooRi                              & 1st  & 96.5\% \\
& AI\_TLfanclub                      & 8th  & 70.5\% \\
& \textbf{DS@GT --- FinMMEval Task 1} & \textbf{9th}  & \textbf{64.5\%} \\
& Vitt Manthan                       & 11th & 26.5\% \\
\midrule
\multirow{4}{*}{English}
& PooRi                              & 1st  & 97.5\% \\
& TCLabs                             & 8th  & 79.0\% \\
& \textbf{DS@GT --- FinMMEval Task 1} & \textbf{9th}  & \textbf{69.5\%} \\
& FinEval                            & 13th & 26.5\% \\
\midrule
\multirow{4}{*}{Hindi}
& fosu ltw                           & 1st  & 92.0\% \\
& NLP-DE                             & 6th  & 85.5\% \\
& \textbf{DS@GT --- FinMMEval Task 1} & \textbf{7th}  & \textbf{73.5\%} \\
& Vitt Manthan SVNIT                 & 10th & 31.0\% \\
\bottomrule
\end{tabular}
\end{table}

\section{Discussion}

The results reported in Section~\ref{sec:results} challenge two assumptions that pervade multilingual NLP system design: that a sufficiently capable model can serve as a universal backbone across languages and that more sophisticated reasoning strategies yield uniform gains. Both assumptions fail in the financial domain, and the failure modes are language-specific, sometimes significant, and not predictable from model size or benchmark reputation alone.
 
\subsection*{Model routing as a correctness requirement}
 
The most consequential finding in Table~\ref{tab:model-routing} is not that Qwen3-14B improves on Arabic, Chinese, and Hindi — its stronger multilingual pretraining makes that expected — but that it simultaneously regresses by 22.9 percentage points on English relative to Qwen2.5-14B under identical prompt, retrieval, and scoring conditions. This asymmetry is structural, not noise; it persisted across every ablation rerun. The mirror image appears in Greek, where Llama-3.1-8B, a smaller model, outperforms all Qwen variants by more than 19 points. Capability rankings among LLMs are not globally transitive, and there is no parameter-count proxy that reliably predicts cross-language ordering. A uniform model selection policy would have erased over 20 accuracy points on English and a comparable margin on Greek — losses that no retrieval improvement could have recovered. The routing function in Equation~\ref{eq:routing} is therefore not an engineering convenience; it is a prerequisite for competitive performance.
 
\subsection*{Why RADS outperforms generation in this setting}
 
Direct option scoring has been shown to match or exceed generation-based methods across multiple-choice benchmarks since MMLU ~\cite{hendrycks2021mmlu} and GPT-3 \cite{brown2020gpt3}, but the advantages are especially pronounced in multilingual financial evaluation. Parse reliability is the clearest: RADS achieved a 100\% parse rate across all 8,869 Chinese validation questions, while few-shot generation (Solver A) required a multi-stage regex fallback and still produced ambiguous outputs on approximately 2\% of Arabic queries. In a 200-question test set, that residual failure rate introduces measurable bias into accuracy estimates and leaderboard positions.
 
The more subtle advantage is architectural. Qwen3's default \texttt{<think>...</think>} token chain, emitted before the answer under standard decoding, displaces the answer letter from the next-token position and breaks Equation~\ref{eq:rads} entirely. Disabling thinking mode via \texttt{enable\_thinking=False} was a prerequisite, not an optimization — without it, Arabic RADS accuracy collapsed to near-chance in preliminary runs. This illustrates a broader point: scoring strategy and model configuration are not independent choices. RADS enforces a clean contract between generation infrastructure and answer extraction that generation-based solvers do not.
 
\subsection*{Chain-of-thought is task-type-sensitive, not just language-sensitive}
 
The 69.8 percentage-point collapse in Greek accuracy under CoT with self-consistency (Solver D, Table~\ref{tab:model-routing}) is the sharpest result in the paper and deserves more than a surface reading. Greek Plutus-MultiFin is a text classification task: questions require assignment to one of six financial categories, not multi-step quantitative derivation. Chain-of-thought prompting was designed for the latter; applied to the former, it biases the model toward narrative completion rather than categorical commitment, and the model reliably talks itself out of the correct label. A similar, if less dramatic, degradation appears on English MMLU-financial ($-34.3$ pp vs.\ RADS), where quantitative reasoning questions are sensitive to the response-format instability introduced by temperature sampling across five CoT paths.
 
The only positive signal for CoT was Chinese, where preliminary tracking reached approximately 66\% — roughly comparable to Qwen3 RADS but at five times the per-query cost. This cost-performance ratio is difficult to justify operationally. Taken together, these results argue for task-type-aware strategy selection as a design axis alongside language-aware model routing: the question of \emph{how} the model reasons matters as much as \emph{which} model reasons.
 
\subsection*{Retrieval quality versus base-model capacity in low-resource languages}
 
Expanding the Hindi knowledge base from cross-lingual RRF proxies to a dedicated native shard of 5,982 exemplars produced a gain of 1 percentage point (Table~\ref{tab:routing-dev}). The gap between our best local system and Claude Opus 4.7 on the same dev set is 17 points (Table~\ref{tab:hindi-frontier}). These two numbers tell the same story from different angles: the performance bottleneck for Hindi is not retrieval infrastructure but base-model capacity over Hindi financial text. The cross-lingual proxies were already supplying adequate signal; what the system lacked was a model with sufficient pretraining density on the target language and domain. For research teams working under inference-cost constraints, this finding reframes the prioritization question — native exemplar collection for languages at this resource level may yield diminishing returns relative to model selection or targeted fine-tuning.
 
The wider implication concerns the BGE-M3 embedding space itself. That a cross-lingual proxy strategy using English and Arabic exemplars could sustain 78\% Hindi accuracy in KB v1 indicates that the shared multilingual vector space is genuinely functional for zero-shot cross-lingual retrieval in finance, not merely a theoretical property of the model. The RRF trust weights ($w_\ell = 2.0$ for native, $w_\text{global} = 1.0$ for cross-lingual) calibrated this signal appropriately without requiring per-language hyperparameter search.
 
\subsection*{The open-weight performance tier and its implications for evaluation}
 
Our final leaderboard results sit 18.5 to 32.0 percentage points below the top-ranked team across all four evaluated languages. The top three teams consistently exceeded 91\% accuracy — approximately the level we observed for Claude Opus 4.7 on Hindi. We cannot verify which underlying models those teams used, but this performance tier is consistent with frontier API access, and the Hindi ablation demonstrates that retrieval improvements alone cannot close a gap of this magnitude. Our submission used only Qwen3-14B, Qwen2.5-14B, and Llama-3.1-8B, and represents what we believe to be the reproducible open-weight ceiling for this task under academic computing constraints.
 
This creates a practical problem for leaderboard interpretation. When open-weight and API-backed systems are ranked on the same accuracy column without distinguishing access tier, the comparison conflates system design quality with model access. A team that builds a carefully engineered retrieval and routing pipeline with open-weight models is not competing on the same terms as one that wraps a frontier API in a five-shot prompt. Benchmark reporting norms for multilingual financial QA would benefit from explicit tier separation — a point this competition's results make concrete.
 
\subsection*{Summary of design implications}
 
Across all four result sections, a consistent principle emerges: effective multilingual financial QA requires deliberate heterogeneity at every layer. Heterogeneous model selection by language, heterogeneous reasoning strategies by task type, and heterogeneous retrieval routing by resource availability each contribute independently to system performance. No single component subsumes the others, and the interactions between them—as illustrated by the Qwen3 thinking-mode failure, the Greek CoT collapse, and the Hindi retrieval plateau—are not always predictable from component-level properties. Uniform pipelines, regardless of the strength of their individual parts, leave systematic accuracy on the table in multilingual settings.

\section{Future Work}

The current analysis reveals several promising directions for advancing multilingual financial exam question-answering capabilities. Building upon our demonstrated semantic fidelity across languages, future research will pursue three complementary directions.

\textbf{Multimodal Integration}: The evolution toward financial reasoning in a multimodal context represents our primary long-term objective. This development will focus on analytical integration across languages and modalities, challenging the system to efficiently integrate financial reports and news, such as charts and regulatory filings for complex analytical QA, testing the ability to generalize across both linguistic and informational modalities \cite{FinMMEval2026}. 

\textbf{Advanced Cross-lingual Adaptation}: Future work will develop systematic approaches 
to model fine-tuning with incrementally larger datasets combined with quality-supervised 
training methodologies. This includes expanded cross-lingual transfer learning strategies 
to better bridge the performance gap between high-resource languages such as English and 
Chinese and low-resource targets such as Greek and Hindi — where, as demonstrated in 
Section~\ref{sec:results}, base-model capacity rather than retrieval quality is the 
dominant performance ceiling. Concretely, we plan to explore continued pretraining on 
domain-parallel financial corpora and cross-lingual distillation from stronger teacher 
models, with the goal of reducing dependence on language-specific model routing by 
developing a single backbone that generalizes reliably across the financial languages 
evaluated here

\textbf{Enhanced Data Augmentation}: Given the language exemplar imbalance in the dataset, we plan to implement data augmentation techniques to enhance representation across low-resource languages, intending to further expand coverage to additional languages and improve system performance and knowledge acquisition in both multilingual and zero-shot financial QA tasks.

\section{Conclusions}

We presented a multilingual retrieval-augmented pipeline for financial exam
question answering across six languages.
The system combines BGE-M3 embeddings with language-partitioned FAISS
indices, weighted RRF fusion for low-resource settings, and RADS for
deterministic, parse-failure-free answer scoring.
 
Three findings stand out.
First, model routing is not optional: a single LLM applied uniformly across
languages leaves significant accuracy on the table.
Qwen3-14B~\cite{qwen3} gains $+4.2$~pp over Qwen2.5-14B~\cite{qwen25} on
Chinese but loses $-22.9$~pp on English; Llama-3.1-8B~\cite{llama3}
outperforms all Qwen variants on Greek by $+21$~pp despite its smaller
parameter count.
Second, reasoning strategies interact with language in non-obvious ways:
chain-of-thought prompting reduces Greek classification accuracy from
$90.7\%$ to $20.9\%$, and self-consistency decoding ($N{=}5$, $T{=}0.7$)
degrades Hindi by $-19$~pp.
Third, implementation details are critical~---~disabling Qwen3's default
thinking mode is a prerequisite for RADS to function; with it enabled,
Arabic accuracy falls from $71.7\%$ to near-chance.
 
Together, these results argue against uniform approaches to multilingual
financial QA and motivate the language-aware, strategy-aware architecture
described here.
Future directions include data augmentation to address knowledge base language
imbalance, extension to multimodal financial inputs such as charts and
regulatory filings~\cite{FinMMEval2026}, and systematic cross-lingual
fine-tuning for low-resource languages.

\section*{Acknowledgements}

We thank the Data Science at Georgia Tech (DS@GT) CLEF competition group for their support.
This research was supported in part through research cyberinfrastructure resources and services provided by the Partnership for an Advanced Computing Environment (PACE) at the Georgia Institute of Technology, Atlanta, Georgia, USA \cite{PACE}. 

\section*{Declaration on Generative AI}
 During the preparation of this work, the author(s) used Claude Sonnet 4.6 and Grammarly in order to: grammar and spelling check, properly format tables, peer review simulation. After using these tool(s)/service(s), the author(s) reviewed and edited the content as needed and take(s) full responsibility for the publication’s content.

\bibliography{main}

\appendix

\section{Prompt for Retrieval-Augmented Direct Scoring}
\label{app:rads-prompt}

\subsection*{System message}
\begin{tcolorbox}[colback=gray!5,colframe=black!50,boxrule=0.4pt]
You are an expert financial exam solver. You will be shown a set of solved example questions, followed by a new question. Select the single correct answer option. Respond with only the letter of the correct option (e.g.\ A, B, C, or D). Do not explain your answer or write anything else.
\end{tcolorbox}

\subsection*{User message}
\begin{tcolorbox}[colback=gray!5,colframe=black!50,boxrule=0.4pt]
Here are similar solved questions to guide you:

\#\#\# Example 1\\
QUESTION: \{exemplar\_1\_stem\}\\
OPTIONS:\\
(A) \{exemplar\_1\_option\_a\}\\
(B) \{exemplar\_1\_option\_b\}\\
(C) \{exemplar\_1\_option\_c\}\\
(D) \{exemplar\_1\_option\_d\}\\
Answer: \{exemplar\_1\_gold\_letter\}

\#\#\# Example 2\\
QUESTION: \{exemplar\_2\_stem\}\\
OPTIONS:\\
(A) \{exemplar\_2\_option\_a\}\\
(B) \{exemplar\_2\_option\_b\}\\
(C) \{exemplar\_2\_option\_c\}\\
(D) \{exemplar\_2\_option\_d\}\\
Answer: \{exemplar\_2\_gold\_letter\}

\ldots\ (Examples 3 through 5 follow the same structure) \ldots

---

Now answer this question:

QUESTION: \{target\_stem\}\\
OPTIONS:\\
(A) \{target\_option\_a\}\\
(B) \{target\_option\_b\}\\
(C) \{target\_option\_c\}\\
(D) \{target\_option\_d\}

Answer:
\end{tcolorbox}

\end{document}